\documentclass[11pt, a4paper]{article}
\usepackage[utf8]{inputenc}
\usepackage{empheq}
\usepackage{amsfonts}
\usepackage{graphicx}
\usepackage{hyperref}
\usepackage[normalem]{ulem}
\usepackage{epigraph}
\usepackage{bbm}
\usepackage{wrapfig}
\usepackage{lineno}
\usepackage{scalerel,stackengine}
\usepackage{physics}
\usepackage[hscale=0.75,vscale=0.8]{geometry}
\usepackage{amsmath, amsthm, amssymb, mathrsfs}
\usepackage{color}
\usepackage{authblk}
\usepackage[dvipsnames]{xcolor}
\usepackage{enumerate}
\usepackage{mathtools}
\usepackage{subcaption}
\usepackage{physics}

\begin{document}

\title{What Bohmian mechanic says about arrival times of 1D vacuum squeezed states}

\author{Angel Garcia-Chung\thanks{{\tt alechung@xanum.uam.mx}} and Humberto G. Laguna\thanks{{\tt hlag@xanum.uam.mx}}}
\affil{Departamento de Qu\'imica, Universidad Aut\'onoma Metropolitana, San Rafael Atlixco No. 186, Iztapalapa, Ciudad de M\'exico 09340, M\'exico}

\date{\today}

\maketitle

\begin{abstract}
We calculate the time of arrival probability distribution of a quantum particle using the Bohmian formalism. The pilot-wave is given by the wave function of the one dimensional vacuum squeezed state written in the Schr\"odinger representation. To obtain this pilot-wave, we used the unitary representation of the symplectic group in the Hilbert space $L^2(\mathbb{R})$. The solution to the Bohmian equation is a closed expression in time thus allowing for a close expression of the time of arrival distribution. We show the dependence of the time of arrival distribution as a function of the squeezing parameter, the ratio of the detector's position and the proper length of the oscillator and the squeezing phase parameter.
\end{abstract}

\section{Introduction}

Gaussian Boson Sampling (GBS) represents a significant milestone in the development of quantum computing, offering a platform to explore computational problems that are infeasible for classical devices. By exploiting the quantum interference of squeezed light in linear optical networks, GBS enables the sampling of complex probability distributions associated with problems such as graph isomorphism, molecular vibronic spectra, and combinatorial optimization \cite{huh2015boson, arrazola2018using, zhong2020quantum}. However, as quantum technologies like GBS evolve toward practical implementations, it becomes increasingly important to deepen our understanding of quantum dynamical properties, including the time of arrival — a fundamental concept linked to how quantum information propagates and is measured. While standard quantum mechanics lacks a self-adjoint time operator, alternative approaches such as the Bohmian interpretation \cite{durr1992quantum, leavens1990transmission} or the use of positive-operator-valued measures (POVMs) \cite{muga2007time} provide operationally meaningful frameworks to define and compute arrival times. Studying these aspects contributes not only to foundational insights but also to the optimization of quantum measurement protocols and the development of temporally sensitive readout schemes in photonic quantum devices. In this context, the interplay between quantum foundations and specialized platforms like GBS underscores the need for a unified approach that bridges theory and experiment.

In classical mechanics, the Time of Arrival (TOA) refers to the time it takes for a particle to travel between two positions along a well-defined trajectory. For a single particle moving in one dimension\footnote{Throughout this work, we consider only one-dimensional single-particle systems.}, the TOA is determined by inverting the Equation of Motion (EoM) and solving for the time parameter. This inversion yields a classical TOA expression dependent on the particle's initial position and momentum. However, this process requires a careful consideration of the EoM's monotonicity and range. If the EoM has a bounded range, the resulting TOA expression may only be valid for specific initial conditions; otherwise, mathematical inconsistencies arise.

To illustrate, consider the classical harmonic oscillator described by the EoM given by
\begin{align}
x(t) = A(x_0, p_0) \cos(\omega t - \phi(x_0, p_0)), \label{EoM}
\end{align}
where the amplitude $A(x_0, p_0)$ and phase $\phi(x_0,  p_0)$ depend on the initial conditions $x_0$ and $p_0$ as
\begin{align}
A(x_0, p_0) = \sqrt{x^2_0 + \frac{p^2_0}{m^2 \omega^2 }}, \qquad \phi(x_0, p_0) = \arctan\left( \frac{p_0}{m \omega x_0} \right).
\end{align}
The range of the EoM given in Eqn~(\ref{EoM}) is the interval $I_{HO} = [-A(x_0, p_0), +A(x_0, p_0)]$ hence, if we seek for the TOA at the position $x(T) = L$ by inverting Eqn~(\ref{EoM}), then we must have that $L \in I_{HO}$ or otherwise the particle will never reach $L$. Now notice that for a fixed value $L$, not all the initial conditions $(x_0, p_0)$ provide an interval $I_{HO}$ such that $L \in I_{HO}$. Therefore, fixing $L$ divides the set of initial conditions $(x_0, p_0)$ into two subsets: one where $L \in I_{HO}$ and thus allowing for finite TOA and another where $L \notin I_{HO}$ and the TOA is undefined. This conclusion extends to any bounded one-dimensional trajectory governed by an invertible EoM.

This insight has significant implications for Bohmian mechanics \cite{cushing1996bohmian, durr2009bohmian, benseny2014applied} where the TOA probability distribution can be constructed by integrating over initial conditions. Consequently, the splitting of initial conditions into those allowing a finite TOA and those where TOA is undefined will inherently shape the resulting TOA distribution. This work explores how this splitting influences TOA distributions in Bohmian mechanics for the 1D vacuum squeezed state.

In Standard Quantum Mechanics (SQM), particle trajectories are not directly available for TOA calculations. Since the splitting of initial conditions arises from the mathematical properties of EoMs, we can anticipate differences between the TOA predictions in SQM and Bohmian mechanics when analyzing 1D systems with finite-range EoMs. However, it is mathematically possible that a similar splitting could also arise as a consequence of specific operations on the wave function. If this is the case, further analysis is required to identify which wave-function-based TOA formulations in SQM incorporate such a splitting.

Numerous approaches \cite{aharonov1961time, kijowski1974time, muga1995time, delgado1997arrival, delgado1998probability, aharonov1998measurement, delgado1999quantum, muga2000arrival} have been proposed to define the TOA in SQM and yet, a definite expression fully satisfactory is still absent. The most recent proposal  \cite{beau2024time} derives the TOA distribution from the wave function by taking the absolute value of the time derivative of the probability density $\rho_t(L)$ at a fixed position $L$. However, this approach does not account for the range constraints of the EoMs, as discussed earlier. Consequently, for systems with finite-range EoMs (as in our results, Eq. \ref{FSS}), the TOA calculation yields a different expression. Additionally, among the most notable proposals to address the TOA in the SQM formalism (see \cite{kijowski1974time, delgado1997arrival}, as well as \cite{grot1996time}), all encounter  mathematical, conceptual or operational problems, e.g, limitations to specific potentials, reliance on ad hoc regulators, or the use of modified Hamiltonians. However, we hope that as part of future projects, we can delve into the TOA predictions based on these proposals when applied to the one-dimensional vacuum squeezed state for a deeper comparison with our results.

In the case of the Bohmian formalism, many works have deal with the TOA \cite{mckinnon1995distributions, das2019arrival, das2019exotic, drezet2024arrival, ruggenthaler2005times} but so far, none of them consider the 1D vacuum squeezed state as the pilot wave for the Bohm trajectories. We are not going to delve into the benefits and relevance of squeezed states as there is a vast literature in this regard. Instead we will point to the main references \cite{walls1983squeezed, schnabel2017squeezed, gerry2023introductory} and of course, readers can find other references therein. The reason as to why the scarcity of Bohmian-like analysis using squeezed states are, in our own opinion, due to squezeed states are mostly analyzed within the Fock representation rather than in the Schr\"odinger representation. The Fock representation is better suited for quantum optics calculations - photon number is a crucial observable - whereas the Schr\"odinger representation is the appropriate one if we want to consider the Bohmian formalism. In the Bohm formalism the position observable plays a dominant role.

Given these considerations, this work calculates and examines the TOA distribution of the 1D vacuum squeezed state using Bohmian mechanics. We focus on a system which weakly interact with the detector and analyze the key features of the TOA distribution and its mean value. A central result is that Bohmian trajectories obtained share similarities with classical harmonic oscillator trajectories in that both are bounded. To proceed, we employ the unitary representation of the symplectic group \cite{moshinsky1971linear, wolf2016development, torre2005linear, wolf2013integral, chacon2021relation} to obtain a position Schr\"odinger representation of the squeezed state. A direct benefit of this approach is the Gaussian form of the quantum state, which depends on the squeezing parameter and phase. Furthermore, since the quantized system corresponds to a quantum harmonic oscillator, we exploit the fact that its time evolution (at both classical and quantum levels) can be described as a curve in the symplectic group, rendering the results in a closed analytical form. These aspects are discussed in Section \ref{SSdescription}. Section \ref{BohmA} presents the Bohmian analysis of the time-evolved squeezed state, with particular emphasis on the dependence of Bohmian trajectories on the squeezing parameter and phase. The TOA analysis for Bohmian trajectories follows in Section \ref{TOASection}, with a discussion of the results provided in Section \ref{Discusion}.

\section{Schr\"odinger representation for squeezed states} \label{SSdescription}

In this section we are going to introduce the main mathematical ingredients to obtain a closed expression for the vacuum 1D squeezed state in the Schr\"odinger representation. We will use this representation for its ease in yielding the explicit form of the Bohm's equations in the next section. To begin with, let us consider the single mode vacuum squeezed operator given by
\begin{align}
\widehat{S}(\xi) = e^{\frac{1}{2} \left[ \xi^* \left( \widehat{a} \right)^2 - \xi \left( \widehat{a}^\dagger \right)^2 \right] }, \label{SOpe}
\end{align}
where the parameter $\xi \in \mathbb{C}$ is the squeezing parameter and let us denote its polar and Cartesian forms as $ \xi = r e^{i \phi} = \xi_x + i \xi_y $. The operator in the  exponential of Eqn~(\ref{SOpe}) is the infinitesimal squeezing operator $\widehat{s}(\xi)$. It can be written in terms of the position and momentum operators $\widehat{x}$ and $\widehat{p}$ as
\begin{align}
\widehat{s}(\xi) = - \frac{i}{2} \left[ \xi_y \frac{\widehat{x}^2}{l^2} - \xi_x \frac{(\widehat{x} \widehat{p} + \widehat{p} \widehat{x})}{\hbar} - \frac{\xi_y \, l^2 \, \widehat{p}^2}{\hbar^2} \right],
\end{align}
where we have introduced the proper length $ l = \sqrt{\frac{\hbar}{m \, \omega}}$. This was done in order to have the units for the coefficients $\xi_x$ and $\xi_y$ compatible with the units for the symplectic group elements as we will see further below. The operator $\widehat{s}(\xi)$ can be written as
\begin{align}
\widehat{s}(\xi) = - \frac{i}{2\hbar} \widehat{\vec{R}}^T \, {\bf L}(\xi) \, \widehat{\vec{R}},
\end{align}
where the array $\widehat{\vec{R}}$ is given by $ \widehat{\vec{R}}^T = \left( \begin{array}{cc} \widehat{x} & \widehat{p} \end{array}\right) $, the symbol ${^T}$ is the array transpose and the matrix ${\bf L}(\xi)$ is given as
\begin{align}
{\bf L}(\xi) =  \left( \begin{array}{cc} \frac{\hbar \, \xi_y}{l^2} & - \xi_x \\ - \xi_x & - \frac{l^2 \, \xi_y}{\hbar} \end{array}\right).
\end{align}

There exists a Lie algebra isomorphism between operators of the form $\widehat{s}(\vec{r})$ for general real $2 \times 2$ symmetric matrices ${\bf L}(\vec{r})$ and the Lie algebra $sp(2,\mathbb{R})$ of the symplectic group with $n=1$ \cite{chacon2021relation}. The isomorphism is given as 
\begin{align}
\iota: sp(2, \mathbb{R}) \rightarrow {\cal P}(2,\mathbb{R}); {\bf J}{\bf L}(\vec{r}) \mapsto \widehat{s}(\vec{r}) 
\end{align}
\noindent where the matrix ${\bf J}$ is of the form
\begin{align}
{\bf J} = \left( \begin{array}{cc} 0 & 1 \\ -1 & 0 \end{array}\right).
\end{align}
\noindent As a result, we can consider the squeezing operator in Eqn~(\ref{SOpe}) to be the unitary representation of a symplectic matrix ${\bf M}(\xi) \in Sp(2,\mathbb{R})$ given as
\begin{align}
{\bf M} = e^{ {\bf J} {\bf L}} = \cosh(\sqrt{- \det \bf L}) \, {\bf 1}_2 + \frac{\sinh(\sqrt{- \det \bf L}) {\bf J}{\bf L}}{\sqrt{- \det \bf L}},  \label{SymGRel}
\end{align}
\noindent where ${\bf 1}_2$ stands for the $2 \times 2$ identity matrix. Notice that in the case of $\widehat{s}(\xi)$, we have that $ {\det \bf L} (\xi) = - r^2$ hence, the components of the symplectic matrix ${\bf M}(\xi)$ are given as
\begin{align}
A = [{\bf M}(\xi)]_{11} &= \cosh(r) - \sinh(r) \cos(\phi), \\
B = [{\bf M}(\xi)]_{12} &= - \frac{l^2 \, \sinh(r) \, \sin(\phi)}{\hbar}, \\ 
C = [{\bf M}(\xi)]_{21} &= -  \frac{\hbar \, \sinh(r) \, \sin(\phi)}{l^2} ,\\
D = [{\bf M}(\xi)]_{22} &= \cosh(r) + \sinh(r) \cos(\phi),
\end{align}
and we can check the correct units of the components recaling that the matrix ${\bf M}$ acts linearly upon the phase space vectors, i.e., $ {\bf M} \vec{R} = \vec{R}'$ .

We now use the unitary representation of this symplectic group element \cite{moshinsky1971linear, wolf2016development, torre2005linear, wolf2013integral, chacon2021relation} and obtain
\begin{align}
\Psi_\xi(x) = \int \frac{ e^{ \frac{i}{2\hbar} \left[ \frac{D}{B} \, x^2 - \frac{2 \, x' \, x}{B} + \frac{A}{B} \, x'^2 \right]} }{  \sqrt{ 2 \pi i \hbar \, B}   }\Psi_0(x') \, d x', \label{URep}
\end{align}
where the vacuum state $\Psi_0(x)$ is given by
\begin{align}
\Psi_0(x) = \left( \frac{1}{\pi \, l^2} \right)^{1/4} \, e^{- \frac{x^2}{2 \, l^2}}. \label{VState}
\end{align}
The integral in Eqn~(\ref{URep}) can be explicitly solved and it provides the analytic expression for the squeezed state written in the Schr\"odinger representation
\begin{align}
\Psi_\xi(x) = N \, e^{- \frac{1}{2} S x^2 }, \label{SState}
\end{align}
where the coefficients $N$ and $S$ are given by
\begin{align}
N & = \left( \frac{1}{\pi \, l^2} \right)^{1/4} \left( \frac{l^2}{l^2 \, A + i \hbar \, B }\right)^{\frac{1}{2}} , \\
S &= \frac{l^2}{\hbar^2 \, B^2 + A^2 \, l^4} - \frac{i}{\hbar} \frac{( l^4 \, A \, C + \hbar^2 \, B \, D )}{(A^2 \, l^4 + \hbar^2 \, B^2)},
\end{align}
respectively.  Here we made use of the symplectic group condition 
\begin{align}
{\bf J} = {\bf M} \, {\bf J} \, {\bf M}^T
\end{align}
which in this case implies that ${\det \bf M} = 1$. Note also that both, $\Psi_0$ and $\Psi_\xi$ are normalized as $\widehat{S}(\xi)$ is unitary represented.

We now move to the time evolution of the squeezed state given in Eqn~(\ref{SState}). We consider the Hamiltonian for the time evolution to be given by the quantum harmonic oscillator Hamiltonian
\begin{align}
\widehat{H} = \frac{1}{2m} \widehat{p}^2 + \frac{m \, \omega^2}{2} \widehat{q}^2 ,\label{QHamil}
\end{align}
which is used to derive the time evolution operator $\widehat{U}(t) = e^{- \frac{i}{\hbar} t \widehat{H}}$. However, the Hamiltonian in Eqn~(\ref{QHamil}) (for a fixed value ot $t$) is also an element of the Lie algebra $sp(2,\mathbb{R})$ where ${\bf L}(H)$ now is given by
\begin{align}
{\bf L}(H) = \left( \begin{array}{cc} m \omega^2 t & 0 \\ 0 & \frac{t}{m}\end{array}\right),
\end{align}
which, using the relation in Eqn~(\ref{SymGRel}) gives a curve $t \mapsto Sp(2,\mathbb{R})$ in the symplectic group, i.e.,
\begin{align}
{\bf M}_H = \left( \begin{array}{cc} \cos(\omega t) & \frac{l^2}{\hbar} \sin(\omega t) \\ - \frac{\hbar}{l^2} \sin(\omega t) & \cos(\omega t) \end{array}\right).
\end{align}

As a result, the time evolution of the squeezed state $\Psi_\xi(x)$ will obey the following relation
\begin{align}
\Psi_\xi(x,t) = \widehat{U}(t) \Psi_\xi(x,t = 0) =  \widehat{S}(H,\xi) \Psi_0(x,t = 0) \label{timeevoState}
\end{align}
where the operator $\widehat{S}(H,\xi)$ is the unitary operator corresponding to the representation of the symplectic matrix
\begin{align}
{\bf M}(t,\xi) = {\bf M}_H \cdot {\bf M}(\xi) = \left( \begin{array}{cc} \widetilde{A} & \widetilde{B} \\ \widetilde{C} & \widetilde{D} \end{array}\right) 
\end{align}

This route of analysis paves the way to obtain the explicit analytic expression for the time evolved squeezed state which takes a form similar to Eqn~(\ref{SState}) but with the matrix coefficients now replaced by those of the matrix ${\bf M}(t,\xi)$ instead. That is to say, $\Psi_\xi(x,t)$ is given as
\begin{align}
\Psi_\xi(x,t) =  \widetilde{N} \, e^{- \frac{1}{2} \widetilde{S} x^2 } \label{FSS}
\end{align}
where the coefficients are now given by
\begin{align}
\widetilde{N} &= \left( \frac{1}{\pi \, l^2} \right)^{1/4} \left( \frac{l^2}{l^2 \, \widetilde{A} + i \hbar \, \widetilde{B} }\right)^{\frac{1}{2}} \\
\widetilde{S} &= \frac{l^2}{\hbar^2 \, \widetilde{B}^2 + \widetilde{A}^2 \, l^4} - \frac{i}{\hbar} \frac{( l^4 \, \widetilde{A} \, \widetilde{C} + \hbar^2 \, \widetilde{B} \, \widetilde{D} )}{( \widetilde{A}^2 \, l^4 + \hbar^2 \, \widetilde{B}^2)} .
\end{align}

\begin{figure}[h!] 
     \centering
     \begin{subfigure}[h]{0.48\textwidth}
         \centering
         \includegraphics[width=\textwidth]{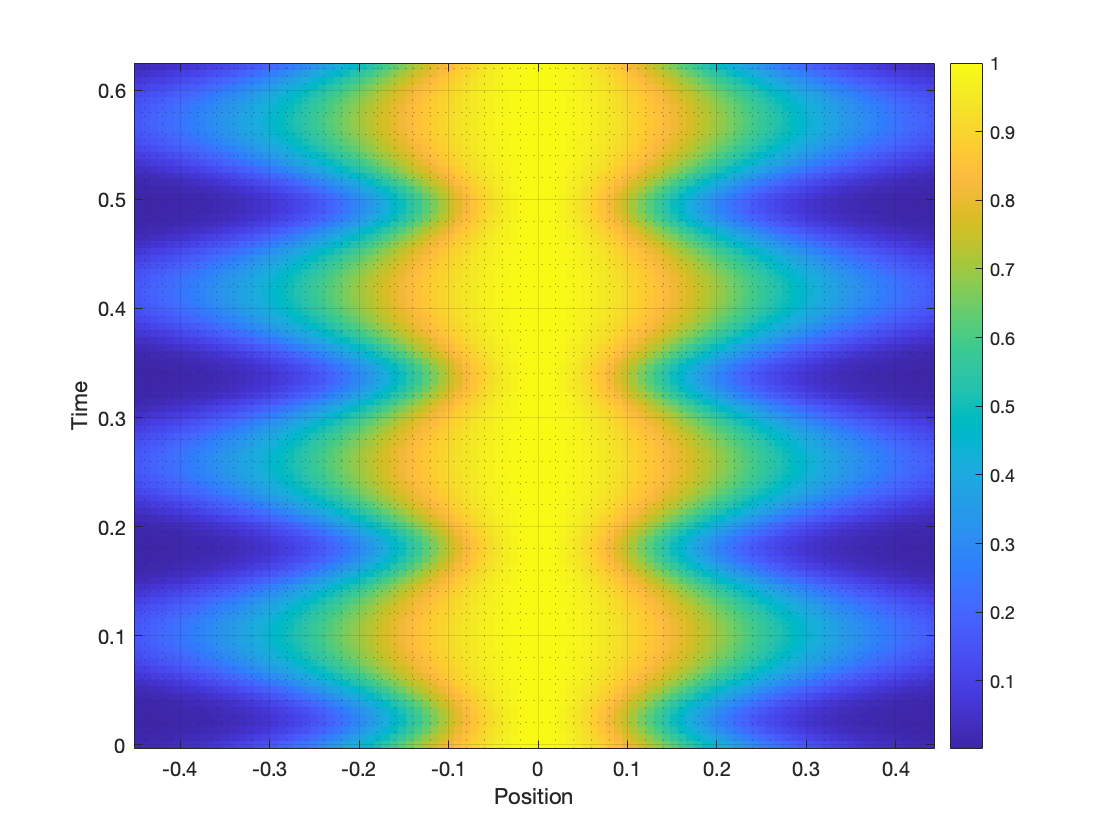}
         \caption{}
         \label{HeatMap0.5}
     \end{subfigure}
     \begin{subfigure}[h]{0.48\textwidth}
         \centering
         \includegraphics[width=\textwidth]{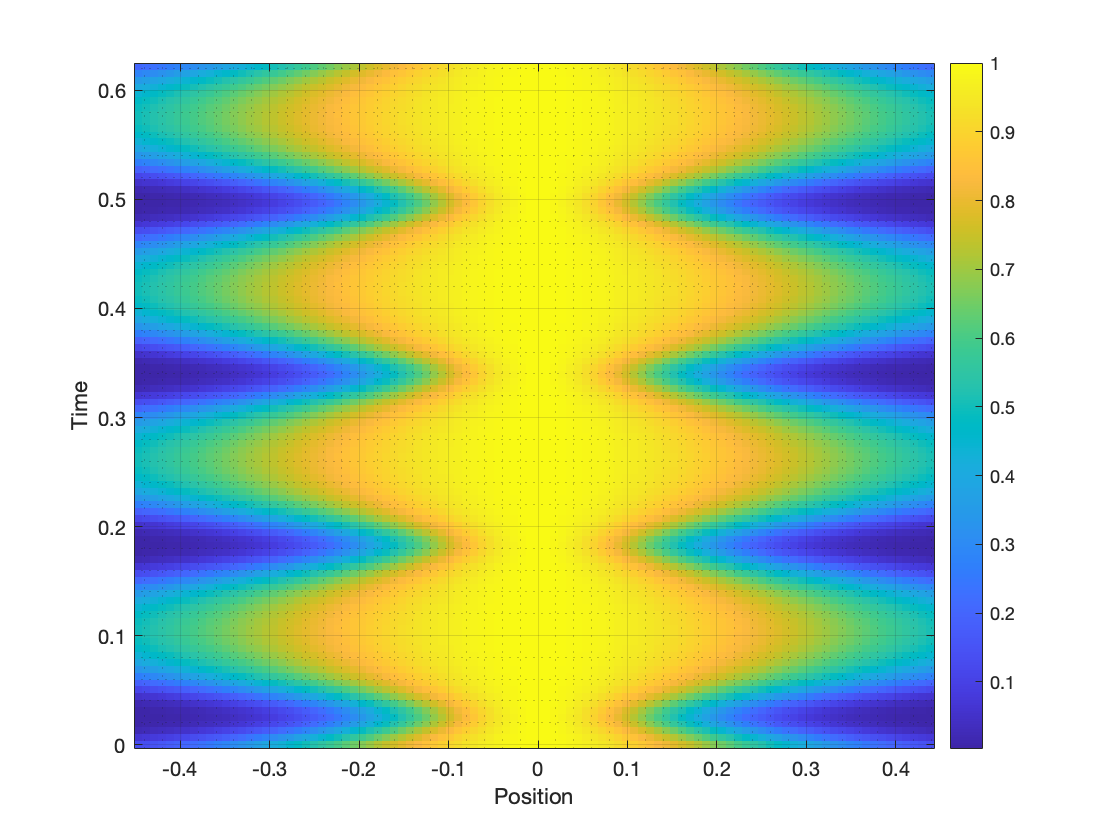}
         \caption{}
         \label{HeatMap1}
     \end{subfigure}
        \caption{{\footnotesize{Heat map of the function $| \Psi_\xi(x,t)|^2$. In (\ref{HeatMap0.5}) $r = 0.5$  and $r = 1$ in (\ref{HeatMap1}), illustrating the effect of the squeezing parameter $r$, both cases with $\phi = 60^\circ$. }}}
        \label{WaveFunction}
\end{figure}

The expression for $\Psi_\xi(x,t)$ is a closed analytic expression which can be used to obtain the exact solution of the Bohm equation for a particle in a Bohm potential guided by the squeezed state $\Psi_\xi(x,t)$. In Figure~(\ref{WaveFunction}) we show a colorbar plot of the square of the wave function $|\Psi_\xi(x,t)|^2$.
It can be notice that those points located close to the axis, have a higher probability and the effect of the squeezing parameter in shrinking some regions whereas it expands others.  There is abundant literature on the properties and main features of this state , e.g., \cite{walls1983squeezed, schnabel2017squeezed, gerry2023introductory}. Let us now move towards the next section and obtain Bohm's equation and its solution. 


\section{Bohmian analysis} \label{BohmA}

\begin{figure}[h!] 
     \centering
     \begin{subfigure}[h]{0.45\textwidth}
         \centering
         \includegraphics[width=\textwidth]{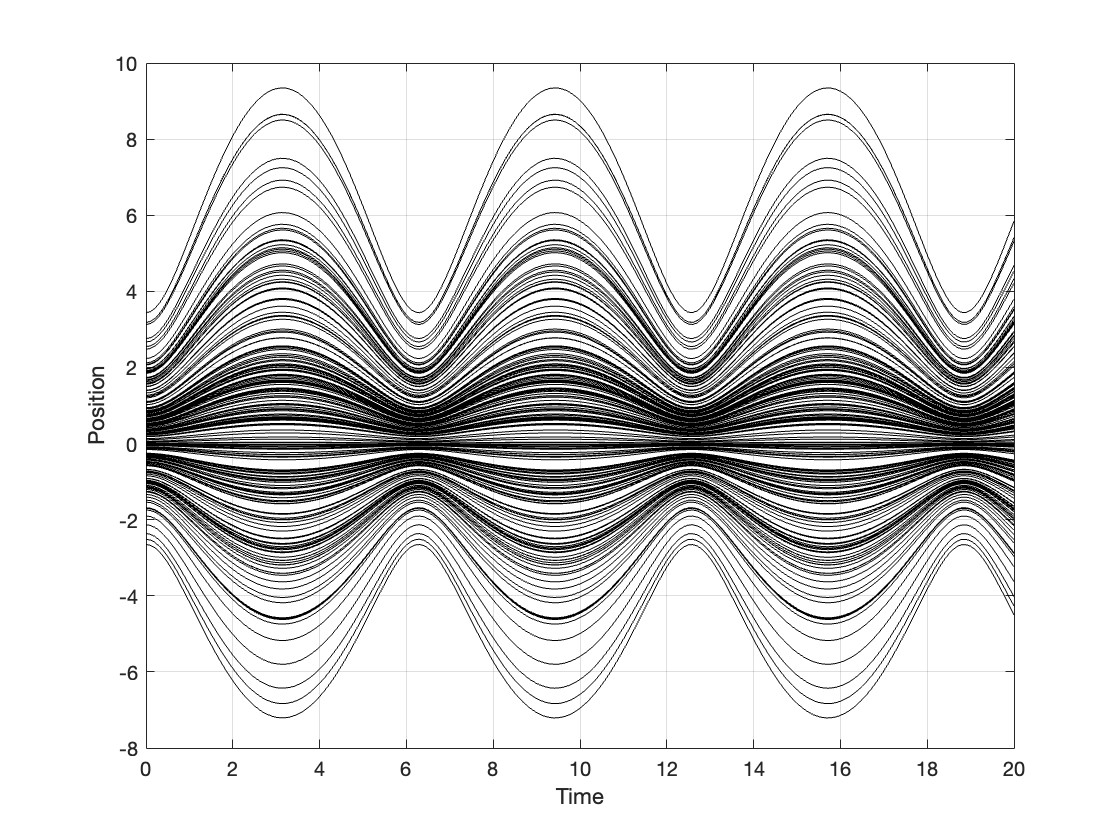}
         \caption{}
     \end{subfigure}
     \hfill
     \begin{subfigure}[h]{0.45\textwidth}
         \centering
         \includegraphics[width=\textwidth]{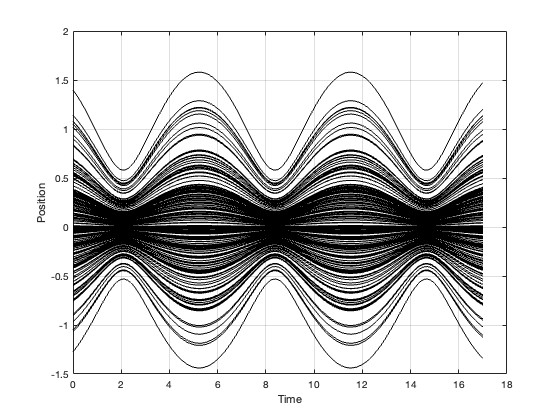}
         \caption{}
     \end{subfigure}
        \caption{{\footnotesize{Bohmian trajectories of the squeezed state in Eqn~(\ref{FSS}) using the squared absolute value of the wave function in Eqn~(\ref{SState}) for the probability distribution of the initial conditions. In (a) the phase $\phi = 0$  and in (b) the phase $\phi = 2 \pi/3$. In both cases we plotted $n = 200$ trajectories with $\hbar = m = 1$ and $\omega = 0.5$ and a squeezed parameter value of $r = 0.5$.}}}
        \label{BT}
\end{figure}

In this section we are going to introduce the analysis of the particles' trajectories using the Bohmian equations  \cite{cushing1996bohmian, durr2009bohmian, benseny2014applied} and then we are going to analyze a scenario in which a position detection is performed. Let us recall that in Bohm's description, the quantum particles have well defined trajectories and velocities. For a single particle in 1D, the Bohm equation is given as
\begin{align}
m \frac{dq}{dt} = - \hbar \, \mbox{Im}\left( \frac{\partial_x \Psi(x,t)}{\Psi(x,t)} \right)\bigg|_{x = q} , \label{BEquation}
\end{align}
where $\Psi(x,t)$ is the wave function or the pilot wave in the Bohmian mechanics context. In our case, replacing the pilot wave by the time evolved squeezed state in Eqn~(\ref{FSS}) gives the following equation 
\begin{align}
\dot{q}(t) = \frac{ \omega \, \tanh(2r) \sin(2\omega t - \phi)}{1 - \tanh(2r) \cos(2\omega t - \phi) } q(t),
\end{align}
whose solution is of the form
\begin{align}
q(t) = q_0 \sqrt{ \frac{1 - \tanh(2r) \cos (2 \omega t - \phi)}{1 - \tanh(2r) \cos \phi }}.  \label{BS}
\end{align}
Here $q_0$ is the initial condition and notice that the period of the trajectories doubles the period of the classical harmonic oscillator system used in the time evolution of the state in Eqn~(\ref{timeevoState}). In Figure~(\ref{BT}) we plotted some of the trajectories using this solution and with initial conditions distributed according to the initial state in Eqn~(\ref{SState}). The figure shows the resemblance with the plot of the wave function given in Figure~(\ref{WaveFunction}).

Using the state in Eqn~(\ref{SState}) for the distribution of the initial conditions is what is called the Bohm thermal equilibrium postulate in which it is stated that not only the wave function is used to derive the Bohm's equations as in Eqn~(\ref{BEquation}) but also that the initial conditions must obey a probability distribution which is not uniform but given by the absolute value squared of the initial state.

Apart from the fact that the trajectories do not cross or touch each other, there are some features of this solution worth to be mentioned. Regarding the initial conditions, (i) If $q_0 = 0$, then $q(t) = 0$ for any $t \geq 0$ and viceversa, having $q(t) = 0$ for some $t$ implies that $q_0 = 0$. (ii) If $q_0 > 0$(or $q_0 < 0$) then $q(t) > 0$(or $q(t) < 0$) for any $t \geq 0$, that is, if we have positive initial conditions then the positions will be positive positions, and viceversa, if we have positive positions they were generated only by positive initial conditions. Similarly for the negative positions counterpart. (iii)  The initial conditions squeeze the trajectories upwards or downwards and the period is constant in all the cases: $T = \pi / \omega$.

Regarding the values of the squeezing parameter $r$, (i) for a given squeezing parameter $r > 0$, the trajectories are squeezed, mostly at those time intervals in which the function $q(t)$ reaches its minimum and expanded on the maximum values. (ii) When $r \rightarrow 0$, we obtain that
\begin{align}
\lim_{r \rightarrow 0} q(t) = q_0,
\end{align}
which indicates that the particle {\it is not moving} and its position remains at the initial position where the particle was emitted. The reason for this is that this limit corresponds to the vacuum state in Eqn~(\ref{VState}) where the particle has a well defined energy $E_0 = \frac{\hbar \omega}{2}$. This implies that the time evolved wave function is a product of two functions, a time dependent phase $e^{ - \frac{i E_0 t}{\hbar}}$ and the position dependent initial quantum state $\Psi_0(x)$. Hence the gradient term in Eqn~ (\ref{BEquation}) leaves the time dependent function unaltered thus cancelling the imaginary contributions. As a result, once we carry out an experiment and measure a given value of the position, we are just detecting the initial position in which the particle was emitted. This is a standard result when considering energy eigenstates in the Bohmian formalism.  (iii) In the case where $r \rightarrow \infty$, we obtain that
\begin{align}
\lim_{r \rightarrow \infty} q(t) = q_0 \bigg| \frac{\sin\left( \omega t - \phi/2 \right)}{ \sin(\phi /2) } \bigg|,
\end{align}
which becomes singular at $\phi = 0$ and $ t \neq n \pi + \phi/2$. 

As mentioned before, the thermal equilibrium hypothesis in Bohm's formalism indicates that the initial positions are distributed according to the initial state given, in this case, by the Eqn~(\ref{SState}). Therefore, once a particle is detected, we cannot predict its initial condition and due to the initial conditions cannot be predicted, we cannot use the trajectories to predict where the particle will be anytime. In this regard, both descriptions of quantum mechanics, the Bohmian and the standard interpretations coincide. 

Let us consider now that a particle is emitted at some unknown initial position $q_0$ and the parameters $r$, $\phi$, $l$ are fixed. According to the Eqn~(\ref{BS}), the maximum and the minimum positions the particle can reach for a given, albeit unknown, initial condition are
 \begin{align}
 q_{max} = q_0 \sqrt{ \frac{1 + \tanh(2r)}{ 1 - \tanh(2r) \cos(\phi)} } , \qquad  q_{min} &= q_0 \sqrt{ \frac{1 - \tanh(2r)}{ 1 - \tanh(2r) \cos(\phi)} },
 \end{align}
therefore, due to $q_0$ cannot be predicted the maximum and the minimum values cannot be predicted either. Of course, the initial condition $q_0$ most be such that $q_{min} \leq q_0 \leq q_{max}$. This condition is always fulfilled due to
\begin{align}
\sqrt{1 - \tanh(2r)} \leq \sqrt{1 - \tanh(2r) \cos(\phi)} \leq \sqrt{1 + \tanh(2r)},
\end{align}
for any $r \geq 0$ and $\phi \in [0, 2 \pi)$. 

\begin{figure}[h!] 
     \centering
         \begin{subfigure}[h]{0.48\textwidth}
         \centering
         \includegraphics[width=\textwidth]{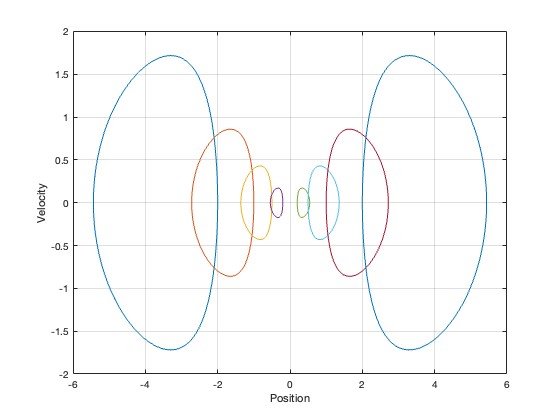}
         \caption{}
         \label{SinPStrajectories}
     \end{subfigure}
     \hspace{-0.5cm}
         \begin{subfigure}[h]{0.48\textwidth}
         \centering
         \includegraphics[width=\textwidth]{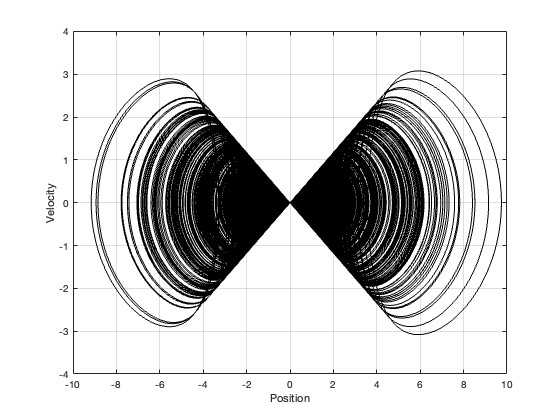}
         \caption{}
         \label{SinPStrajectories2}
     \end{subfigure}     
     
        \caption{{\footnotesize{Bohmian trajectories of the squeezed state in Eqn~(\ref{FSS}) using the squared absolute value of the wave function in Eqn~(\ref{SState}) for the probability distribution of the initial conditions. In (a) the phase $\phi = 0$  and in (b) the phase $\phi = 2 \pi/3$. In the second case we plotted $n = 200$ trajectories and in both we have: $\hbar = m = 1$, $\omega = 0.5$ and a squeezed parameter value of $r = 0.5$. The purpose of (a) is to show the shape of individual trajectories whereas in (b) the collective shape of the trajectories.}}}
        \label{PhaseSpace}
\end{figure}
In Figure~(\ref{SinPStrajectories}) we plotted the trajectories in the phase space $(q(t), \dot{q}(t))$ for different values of the initial conditions. We can notice in Figure~(\ref{SinPStrajectories2}) the effect of having larger values of initial conditions upon the trajectories: the largest the initial condition, the largest the maximum and the lowest minimum velocities of the system as well. Actually, we found a forbidden region in the phase space delimited by the lines
\begin{align}
\dot{q}_{\pm} = \pm 2 \, \omega \, \sinh(2r) \, q.
\end{align}
We are intrigued about the role of this region when considering the multiple Fourier modes of the quantum free field (for example a real free scalar field). In particular, its role when considering the Lorentz symmetry. However, such analysis is beyond the scope of our present work. 

Up to this point, notice the similarity between the classical and the quantum Bohmian description in the sense that both give rise to closed trajectories in the phase space and therefore, the positions (as well as the velocities) are bounded. This {\it bounded-ness} of the Bohmian trajectories influences the TOA of the particles for example, it implies that the TOA will have an upper bound. Let us explore all these aspects in the next subsection.

\subsection{Detection at a given position $L$}

\begin{figure}[h!] 
     \centering
     \begin{subfigure}[h]{0.45\textwidth}
         \centering
         \includegraphics[width=\textwidth]{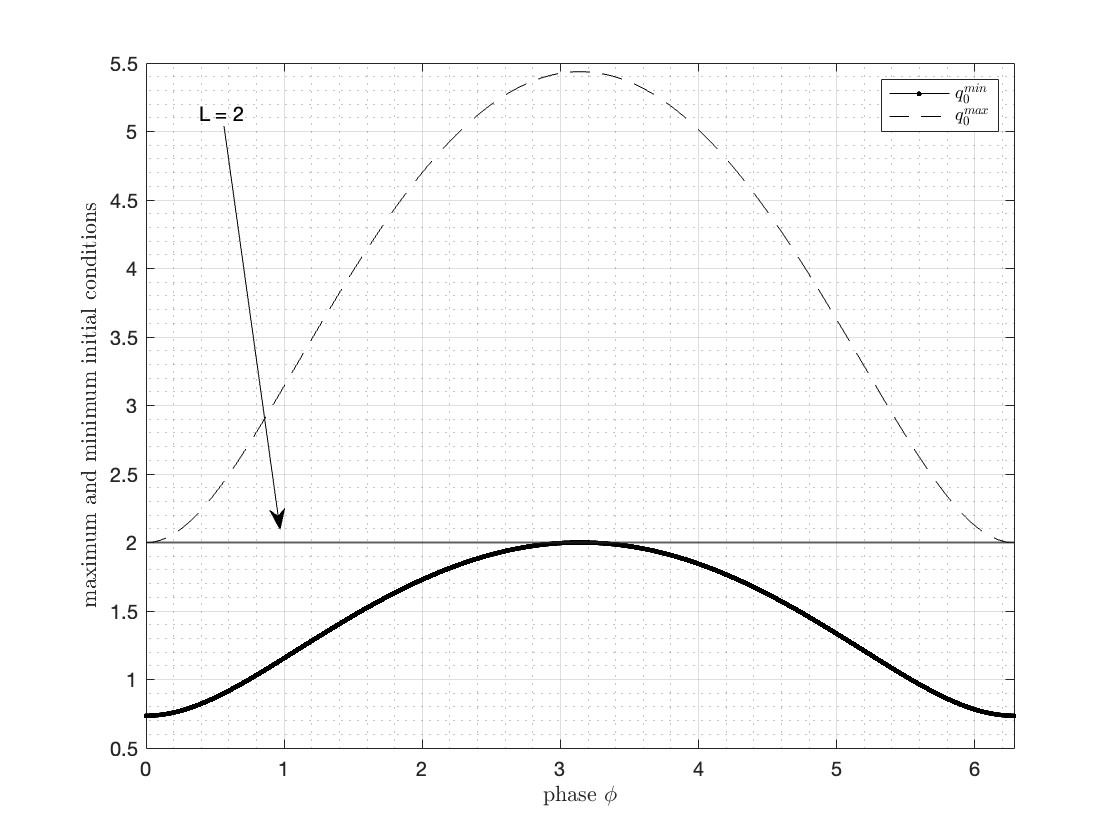}
         \caption{}
     \end{subfigure}
     \hfill
     \begin{subfigure}[h]{0.45\textwidth}
         \centering
         \includegraphics[width=\textwidth]{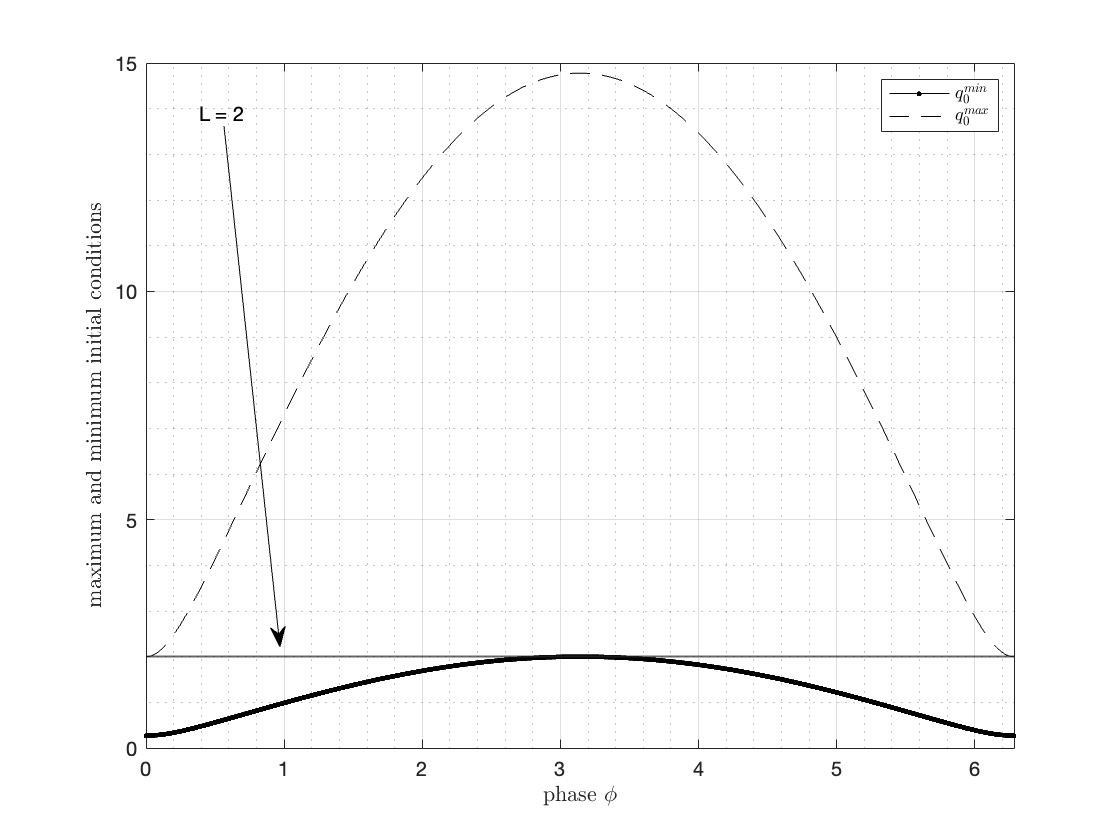}
         \caption{}
     \end{subfigure}
        \caption{{\footnotesize{ Maximum initial condition (dashed line) and minimum initial conditions (solid line) as functions of the phase $\phi$, see Eqn~(\ref{qMin}) and Eqn~(\ref{qMax}). In (a) the squeezing parameter $r  = 0.5$ and in (b) $r = 1$ and in both cases $L =2$. This plot shows that all the trajectories detected at $L = 2$ will have initial conditions lying between these two curves for any $\phi$.}}}
        \label{INiCond}
\end{figure}

Consider the scenario in which the Bohmian particle was detected at some position $L > 0$ (this happens of course at finite time). We do not know what the initial condition was, but we do know that the initial condition associated with the trajectory followed by the particle is in the subset given by those initial conditions such that $L = q(t_{oa},q_0)$ for some finite time $t_{oa}$. From Eqn~(\ref{BS}) it can be checked that the minimum value of the initial condition, denoted by $q^{min}_0$, such that the maximum position of its trajectory is $L$ is given by
\begin{align}
q^{min}_0(r,\phi, L) = L \sqrt{ \frac{ 1 - \tanh(2r) \cos(\phi)}{1 + \tanh(2r)} },  \label{qMin}
\end{align}
and similarly, the maximum value of the initial condition, denoted by $q^{max}_0$, such that the minimum position of its trajectory is $L$ is given by
\begin{align}
q^{max}_0(r,\phi, L) = L \sqrt{ \frac{ 1 - \tanh(2r) \cos(\phi)}{1 - \tanh(2r)} }. \label{qMax}
\end{align}
Hence, if the particle was detected at position $L$, despite not knowing the initial condition, the Bohmian formalism predicts that the initial condition belongs to the interval 
\begin{align}
q_0 \in I_{BSS} = [q^{min}_0(r,\phi, L), q^{max}_0(r,\phi, L)], \label{PosInterval}
\end{align}
\noindent see Figure (\ref{INiCond}) for some illustration. Note that due to both $q^{min}_0$ and $q^{max}_0$ linearly depend on $L$, the interval length is also a linear function of $L$, hence when $L$ increases the length of the interval linearly increases and therefore, {\it more} initial conditions {\it can participate} in the detection events.

Additionally, there is a value of the phase, denoted as $\phi_c(r)$, in which the length of the interval formed with the initial conditions above $L$ equals the length of the interval formed with those initial conditions below $L$, i.e., $q^{max}_0(r,\phi_c, L) - L = L - q^{min}_0(r,\phi_c, L)$. This phase value is found as a solution of the relation
\begin{align}
\cos(\phi_c(r)) = \frac{\left( 2 \cosh^2(r) + 1 \right) \left( \cosh(r) - 1 \right)}{\sinh(2r) \, \cosh^2(r)}.
\end{align}
This shows that the initial conditions values are distributed differently around the detection position $L$. When $ \phi < \phi_c(r)$, then $q^{max}_0 - L < L - q^{min}_0$ and when $ \phi > \phi_c(r)$, then $q^{max}_0 - L > L - q^{min}_0$. On the other hand, from Eqn~(\ref{qMin}) and Eqn~(\ref{qMax}), we notice that
\begin{align}
\lim_{r \rightarrow 0} q^{min}_0(r,\phi,L) = \lim_{r \rightarrow 0} q^{max}_0(r,\phi,L) = L, \label{SLIni}
\end{align}
\noindent indicating that in this limit the particle will be detected at $L$ if and only if it was emitted exactly at $q_0 = L$.  However, we will see further below that this limit is a singular one. Moreover, when $r \rightarrow \infty $ we have that
\begin{align}
\lim_{r \rightarrow \infty} q^{min}_0(r, \phi,L) = L |\sin(\phi/2)|, \qquad \lim_{r \rightarrow \infty} q^{max}_0(r, \phi,L) = \left\{ \begin{array}{ccc} L & \mbox{if } & \phi = 0 \\ \infty & \mbox{if } & \phi \neq 0 \end{array}\right. ,
\end{align}
\noindent which shows that taking the phase value $\phi = 0$ and then the limit ${r \rightarrow \infty}$ yields a result different to the one obtained when taking the limit first and then  evaluating $\phi = 0$. Moreover, when $\phi = 0$ the interval in Eqn (\ref{PosInterval}) is finite while it ranges from a finite value at $L |\sin(\phi/2)|$ to infinity when $\phi \neq 0$.


\section{Time of arrival} \label{TOASection}

We now state the following question: what time would it take for the particle to reach this position $L$ for the first time? Let us call this {\it time of arrival}\footnote{We do not explore {\it how} this time measurement is performed. We are just assuming it can be done and if so, what will be the consequences in light of the Bohmian interpretation. } and let us denote it as $t_{oa}$. We can obtain this time by inverting the function in Eqn~(\ref{BS}) and considering the principal branch of the $\arccos$ function
\begin{align}
t_{oa}(q_0)  = \frac{1}{2 \omega} \left\{ \phi +  \arccos \left[ \frac{1}{\tanh(2r)} \left\{ 1 - \left[ 1 - \tanh(2r) \cos(\phi)\right] \frac{L^2}{q^2_0} \right\} \right] \right\}, \label{BMTime}
\end{align}
and this holds as long as $q_0 \in [q^{min}_0 , q^{max}_0]$, which is the domain of $t_{oa}(q_0)$. Let us analyse $t_{oa}$ as a function of the initial condition $q_0$. If we evaluate $t_{oa}$ at the initial positions given in Eqn~(\ref{qMin}) and Eqn~(\ref{qMax}) in Eqn~(\ref{BMTime}) we obtain
\begin{align}
t_{oa}(q^{min}_0) = \frac{ \phi + \pi}{ 2\omega} , \qquad t_{oa}(q^{max}_0) = \frac{ \phi}{ 2\omega}, \label{TimeValues}
\end{align}
\noindent which indicates that when the $\phi < \pi$ then the maximum time for a particle to reach $L$ is $t_{oa}(q^{min}_0)$ and when the phase $\phi > \pi$ then the maximum time for a particle to reach $L$ is $t_{oa}(q^{max}_0)$. Consequently, no matter how  long $L > 0$ is, as long as there is a detection, the maximum time taken for a particle to reach this position is upper bounded. In both cases, the minimum time is always $t_{oa} = 0$. 

An important aspect of these time values is that they do not depend on the parameter $r$. As a consequence, the limits ${r \rightarrow 0}$ and ${r \rightarrow \infty}$ do not alter these values. However, if we consider the limit ${r \rightarrow 0}$, then from Eqn~(\ref{SLIni}) we know that the initial conditions tend to $L$ indicating that there is only one possible time of arrival given by $t_{oa} = 0$, as there will be only one particle emitted at $L$. But on the other hand, from Eqn~(\ref{TimeValues}) we notice that this is not the case. As a result, we have to consider the limit ${r \rightarrow 0}$ as a singular one for the function in Eqn~(\ref{BMTime}). Of course, this singular behavior results from the inverse function used to solve the parameter $t$ from the Eqn~(\ref{BS}) and the factor of $\tanh(2r)$ in the denominator so nothing of real physical relevance is happening here. The lesson is that when considering the function in Eqn~(\ref{BMTime}), we have to take the limit ${r \rightarrow 0}$ as a singular point\footnote{The singularity emerges due to in the limit $r \rightarrow 0$ the inverse function $\arccos$ becomes multivalued.}.

An observation worth to consider is that due to the domain of the function in Eqn (\ref{TimeValues}) is the interval in Eqn (\ref{PosInterval}), no time value can be derived for initial conditions out of this interval. What can we formally do to consider those initial conditions $q_0 \notin I_{BSS}$? What we consider a natural option to answer this is to construct the time function as a piecewise function of the form
\begin{align}
t_{fa}(q_0) = \left\{ \begin{array}{ccc} \frac{ \phi +  \arccos \left[ \frac{1}{\tanh(2r)} \left\{ 1 - \left[ 1 - \tanh(2r) \cos(\phi)\right] \frac{L^2}{q^2_0} \right\} \right] }{2 \omega}  & \mbox{if} & q_0 \in I_{BSS}, \\ \infty & \mbox{if} & q_0 \notin I_{BSS}, \end{array} \right.
\end{align}
hence indicating that the particle will never reach the position $L$ if $q_0 \notin I_{BSS}$. However, as we will see further below, this brings a scenario in which the probability for these time values becomes problematic and some additional constraints are needed. Recall that, we cannot consider as null the time values for the initial positions out of the interval $I_{BSS}$ due to this implies that the particles indeed reached the position $L$ because they were emitted exactly at $L$ in a similar way to what we have explained when $r = 0$. The physical situation here is rather different and that is why we have to consider this approach.

Let us now continue with the analysis and recall that up to this point, we have only considered one detection event, so let us consider that we repeat this process many times. Each experiment performed with the same values of $L$, $r$ and $\phi$ and each time we collect the values of the time of arrival $t_{oa}$ of the particles. Our questions are now the following: (i) what would be the probability distribution of these time of arrivals? and (ii) what is the mean value of the times collected? Let us address these questions in the following subsection.

\subsection{Time of arrival statistics}

\begin{figure}[h!] 
     \centering
     \begin{subfigure}[h]{0.45\textwidth}
         \centering
         \includegraphics[width=\textwidth]{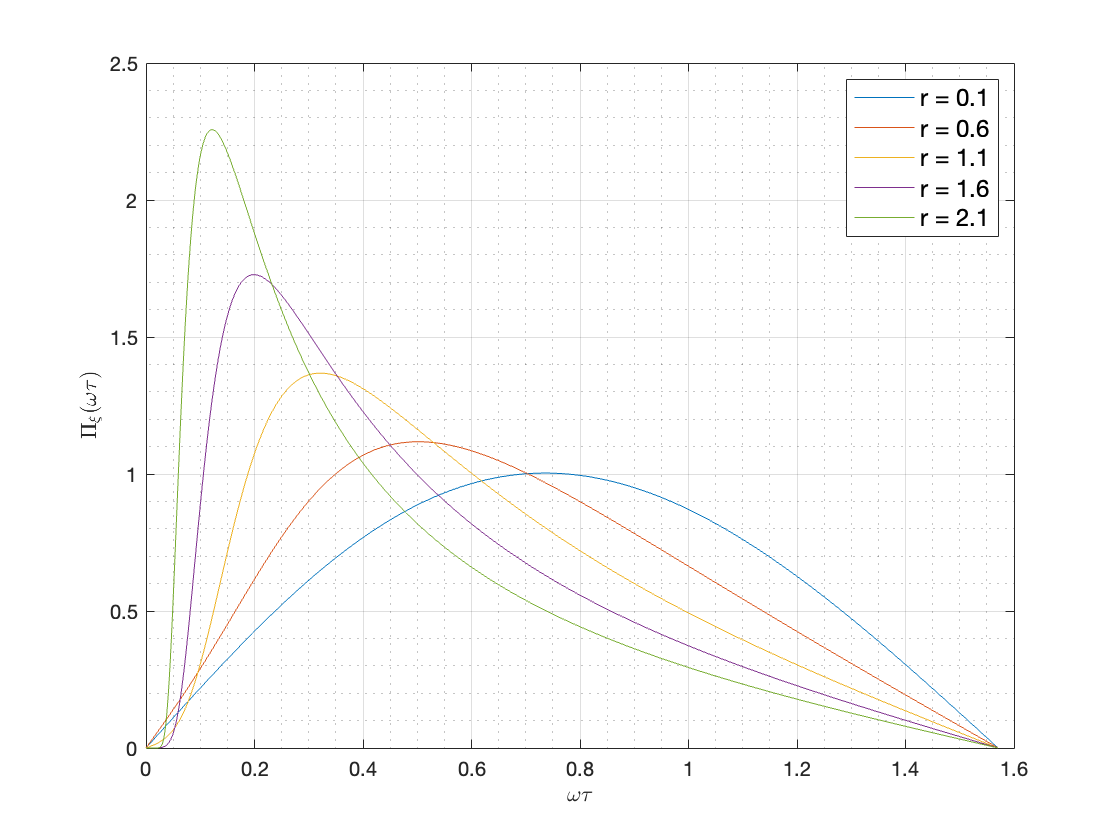}
         \caption{}
         \label{norma}
     \end{subfigure}
     \hfill
     \begin{subfigure}[h]{0.45\textwidth}
         \centering
         \includegraphics[width=\textwidth]{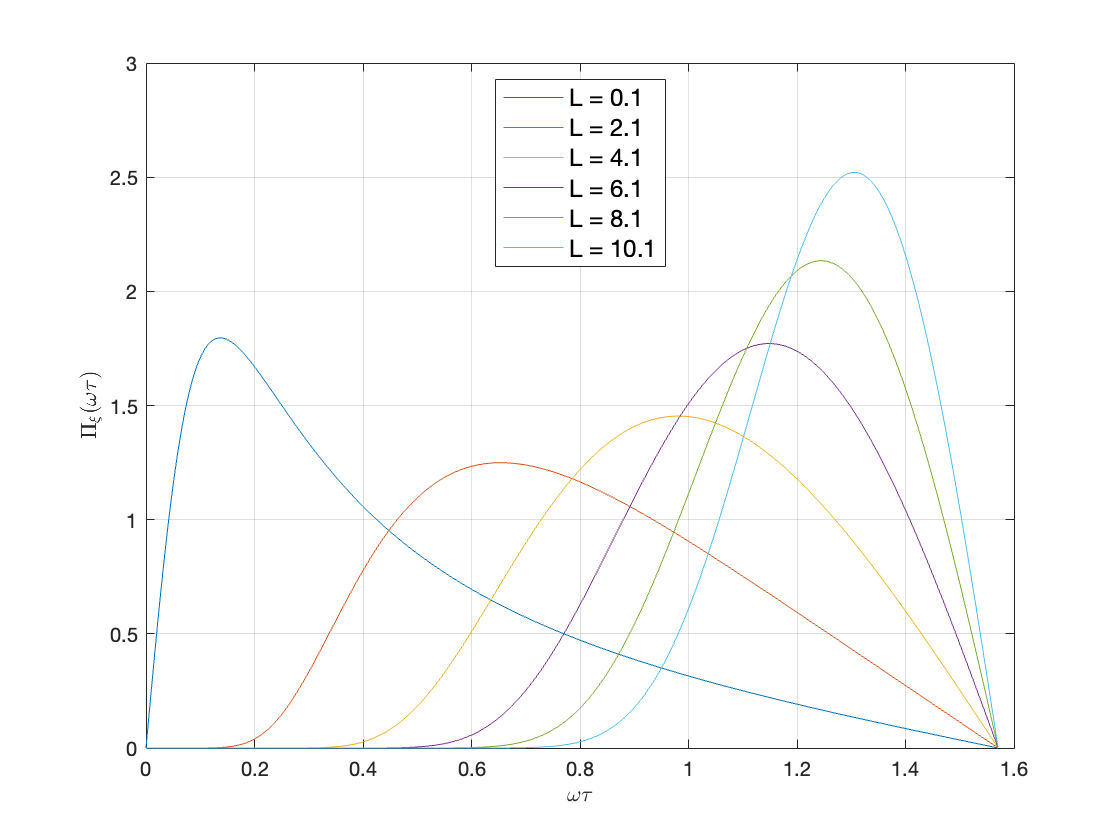}
         \caption{}
         \label{normb}
     \end{subfigure}
        \caption{{\footnotesize{ Probability distribution of Time of Arrival $ \Pi_\xi$ as a function of $\omega \tau$. In ~(\ref{norma}) we varied the squeezing parameter $r$ and consider $L = 1$ fixed. In ~(\ref{normb}) we fixed $r = 1$ and $L$ changes. In both cases, $\hbar = m = 1$, $\omega = 0.5$ and $\phi = 0$.}}}
        \label{ProbTime}
\end{figure}

The probability distribution of the time of arrival has been defined in many references (see for example \cite{mckinnon1995distributions, das2019arrival, das2019exotic, drezet2024arrival, ruggenthaler2005times}) when adapted to the Bohmian description. Denoted as $ \Pi_\xi(t)$ in the present case, its expression is given by
\begin{align}
 \Pi_\xi(\tau) &= \frac {\int_{\mbox{supp} \Psi_\xi} \delta\left( t(q_0) - \tau \right) | \Psi_\xi (q_0) |^2 \, d q_0 }{ Z }, \label{TofAProbDistr}
 \end{align}
 where $t_{min}$ and $t_{max}$ are the extremes of the interval $[\frac{\phi + \pi}{2\omega}, \frac{\phi}{2\omega}]$ according to the values in the Eqn~(\ref{TimeValues}) and the normalization factor $Z$ is given by
 \begin{align}
 Z = \int^{t_{max}}_{t_{min}} \left[ \int_{\mbox{supp} \Psi_\xi} \delta\left( t(q_0) - \tau \right) | \Psi_\xi (q_0) |^2 \, d q_0 \right] dt.
 \end{align}
Eqn (\ref{TofAProbDistr}) expresses the probability distribution of the time of arrival of the particles that will be detected at $L$. Once we substitute the expression for $ \Psi_\xi (q_0)$ and $t(q_0) = \tau$ in Eqn~(\ref{TofAProbDistr}) we obtain the following expression for $ \Pi_\xi(\tau)$
 \begin{align}
 \Pi_\xi(\tau) &= \frac{L \omega \sinh(2 r) \sin(2 \omega \tau - \phi) \,  e^{ - \frac{L^2}{l^2 \left[ \cosh(2r) - \sinh(2r) \cos(2 \omega \tau - \phi)\right]}} }{ Z \sqrt{\pi} \, l \, \left[ \cosh(2r) - \sinh(2r) \cos(2 \omega \tau - \phi)\right]^{3/2}  } \label{TofADistri}
\end{align}
and in Figure~(\ref{ProbTime}) we show its graph as a function of $\omega \tau - \phi$.

\begin{figure}[h!]
\centering
     \begin{subfigure}[h]{0.49\textwidth}
     \centering
     \includegraphics[width=\textwidth]{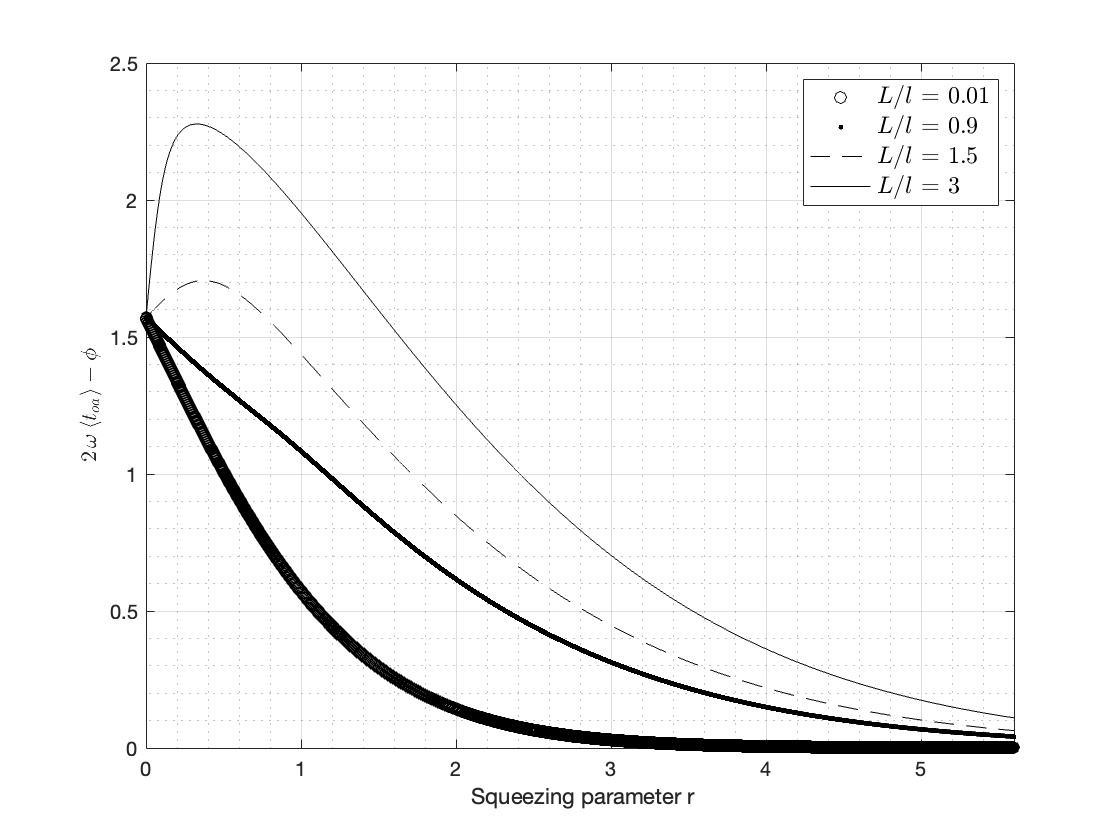}
     \caption{}
     \label{MeanTimePlota}
     \end{subfigure}
     \begin{subfigure}[h]{0.49\textwidth}
     \centering
     \includegraphics[width=\textwidth]{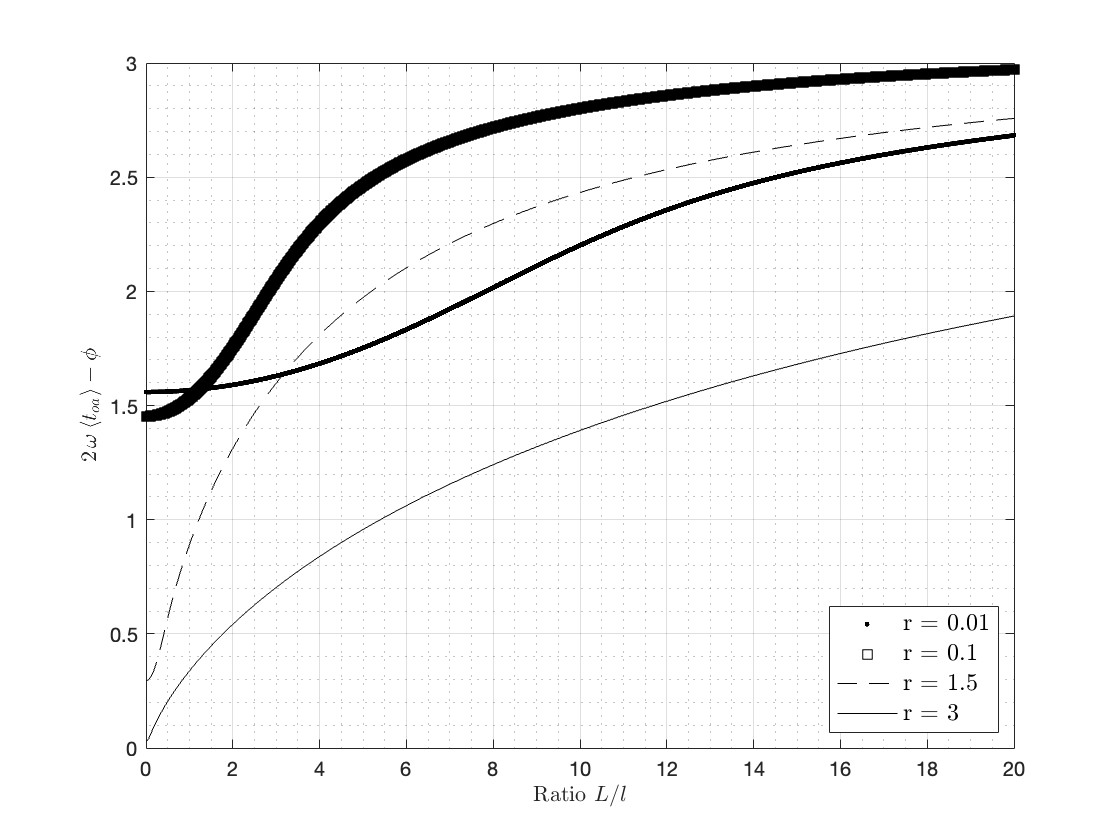}
     \caption{}
     \label{MeanTimePlotb}
     \end{subfigure}
 \caption{{\footnotesize{Plot of $2 \omega \langle t_{oa} \rangle - \phi $ as a function of the squeezing parameter $r$~(\ref{MeanTimePlota}) and the ratio $L/l$~(\ref{MeanTimePlotb}). In ~(\ref{MeanTimePlota}) for different values of the ratio $L/l$ and in ~(\ref{MeanTimePlotb}) for different values of the squeezing parameter $r$.}}}
\label{MeanTimePlot}
\end{figure}

From Figure~(\ref{norma}) we can extract out the following observations, (i) $ \Pi_\xi(\tau)$ has a global maximum value at, say $\tau_{max}$, (ii) increasing $r$ {\it moves} $\tau_{max}$ to the left indicating that the most probable time of arrival decreases with increasing $r$. This result can be observed in Figure~(\ref{MeanTimePlota}) where we showed the values of the mean time of arrival as a function of the squeezing parameter $r$ and for different values of the ratio $L/l$. (iii) The value of $ \Pi_\xi(\tau_{max})$ increases with increasing $r$.  Figure~(\ref{normb}) on the other hand shows a very different behavior.  First, the distribution will always present a global maximum $ \Pi_\xi(\tau_{max})$ for every value of $L$, however, when $L$ increases the value $\tau_{max}$ {\it moves} now to the right. Additionally, when increasing $L$ the values of $ \Pi_\xi(\tau_{max})$ first decrease and then, after a certain value of $L$ it reaches the infimum value among the values of $ \Pi_\xi(\tau_{max})$ and then it starts increasing again.

Let us now introduce the expression for the mean value of the time of arrival using the distribution in Eqn~(\ref{TofAProbDistr})
\begin{align}
\langle t_{fa} \rangle & = \int \tau \, \Pi_\xi(\tau) d\tau,
\end{align}
\noindent and let us re-write this expression in terms of the $q_0$ coordinates using the relation $t(q_0) = \tau$ as
\begin{align}
\langle t_{fa} \rangle & =  \frac { \int^{q^{max}_0(L)}_{q^{min}_0(L)} \left\{ \frac{\phi +  \arccos \left[ \frac{1}{\tanh(2r)} \left\{ 1 - \left[ 1 - \tanh(2r) \cos(\phi)\right] \frac{L^2}{q^2_0} \right\} \right]}{2 \omega} \right\} | \Psi_\xi(q_0)|^2 \, dq_0 }{ \int^{q^{max}_0(L)}_{q^{min}_0(L)}  | \Psi_\xi(q_0)|^2 \, dq_0 },  \label{NORMEDTIME}
\end{align}
where $\Psi_\xi(q_0)$ is the state given in Eqn~(\ref{SState}) but replacing $x$ by $q_0$ and restricting its domain to the integration interval. This expression can be interpreted as the mean value of the function $t(q_0)$ with probability distribution given by
\begin{align}
Pr(Q = q_0) = \left\{ \begin{array}{ccc} \frac{ |\Psi_\xi(q_0)|^2 \, dq_0}{\int^{q^{max}_0(L)}_{q^{min}_0(L)}  | \Psi_\xi(q_0)|^2 \, dq_0} & \mbox{if} & q_0 \in  I_{BSS} \\ 0 & \mbox{if} & q_0 \notin  I_{BSS} \end{array}\right. \label{NewProbDistri}
\end{align}

This result affects our interpretation of the probability for the initial conditions as follows: the initial conditions participating in the detection event at $L$ are not distributed according to Eqn~(\ref{SState}) but according to the distribution given in Eqn~(\ref{NewProbDistri}). The reason for this is because those $q_0 \notin I_{BSS}$ must have a null probability to happen because otherwise, their time of arrival will yield an undefined value. Therefore, only those $q_0 \in I_{BSS}$ have to be considered as physical detection events. As a consequence, the initial condition distributions should be updated which requires the normalization factor in Eqn~(\ref{NewProbDistri}). 

In Bayesian terms, Bohmian mechanics offer more information about the system, which in this case consists in the realization of a position interval for the initial conditions out of which the time taken for the particle to reach the position $L$ is infinite. This {\it new information} about the system drives us to update our information of the probabilities for the different time values to happen as seen in the Eqns~(\ref{TofAProbDistr}) and (\ref{NORMEDTIME}). If, on the other hand, we decide to integrate over the entire real line in Eqn~(\ref{NORMEDTIME}) without changing the values of the probabilities, then a mathematical inconsistency emerges due to imaginary time values associated with the initial positions out of the domain of the time function. This is an immediate conclusion according to our Bohmian results: we are forced to update the probability distribution for the initial conditions or otherwise the mathematical prediction is inconsistent.

In Figure~(\ref{MeanTimePlot}) we plotted the values of the mean time of arrival. In~(\ref{MeanTimePlota}) we plotted the values of the mean TOA as a function of $r$ with different values of the ratio $L/l$ and in~(\ref{MeanTimePlotb}) the means TOA were plotted as a function of the ratio $L/l$ and for different values of $r$. Subfigure (\ref{MeanTimePlota}) shows that when $r$ increases, all the curves tend to the same value of $\phi/2\pi$, constituting a lower bound for $\langle t_{oa} \rangle$. Figure~(\ref{MeanTimePlotb}) on the other hand, shows an upper bound for every fixed value of $r$, when $L/l \rightarrow \infty$ then $ \langle t_{oa} \rangle  \rightarrow  (\phi + \pi)/ 2\pi $. Additionally, for small values  of $r$, Figure~(\ref{MeanTimePlota}) shows that $ \langle t_{oa} \rangle \rightarrow (\frac{\pi}{2} + \phi)/ (2 \omega)$, for any $L/l$ value but as we know, in this limit, the inverse $arccos$ function is not well defined hence it constitutes a singular limit. 

All these limits have in common that they do not depend on the parameters $r$ or $L$. However, the limit of small $L/l$, as can be seen in Figure~(\ref{MeanTimePlotb}) shows that it will depend on the $r$ value fixed.


\section{Discussion and conclusions} \label{Discusion}
 

Let us examine more closely the implications of a finite interval $I_{BSS}$ within the Bohmian framework. In standard quantum mechanics, the position operator has an unbounded spectrum, and consequently, no such intrinsic spatial constraint arises. We can, however, wonder if there is an implications upon the probability distribution for the position aligned with the interpretation in Eqn~(\ref{NewProbDistri}).

 Consider a scenario where a detector is positioned at $L$, and particles are prepared at time $t = 0$ in the quantum state given by Eq. (\ref{SState}), subsequently evolving in time according to Eq. (\ref{FSS}). Within the standard formalism, it is not possible to distinguish among initial positions, as the concept of individual trajectories is absent and the particle is fundamentally delocalized over the entire real line. As a result, shifting the detector’s position modifies the detection statistics solely through changes in the probability density derived from Eq. (\ref{FSS}), now evaluated at the detector's location. There is no notion of a path constraint that governs whether a particle can or cannot reach the detector. In contrast, Bohmian mechanics introduces well-defined trajectories, which can impose natural limits on the set of particles that are dynamically able to arrive at the detector. This contrast highlights how the presence or absence of particle trajectories has observable consequences in the TOA predictions.

 For example, let us consider the following {\it Gedanken experiment}:
\begin{itemize}
\item Let us produce a particle at $t = 0s$ (all the parameters fixed $\lambda = \{ r, \phi, \omega, \dots \}$) described with the state in Eqn~(\ref{SState}). The detector, placed at $L$, is turned on also at $t = 0$ (i.e., particle production and turning on the detector are syncronised events).  
\item As time goes by, the quantum state evolves according to Eqn~(\ref{FSS}) and the particle will hit the detector. We do not know where the particle will hit, neither when, but we can turn the detector off after a time $T = ( \phi + \pi) / (2 \omega)$, with $\phi < \pi$ according to Eqn.~(\ref{TimeValues}). After this, all the detections will be registered the experiment is repeated, say $N$ times. We collect the number of clicks $N^{exp}_c(L, T; \lambda)$ experimentally detected on screen for the same window of time $t \in [0, T]$. 
\item According to the standard interpretation of quantum mechanics, the number of detection events in the interval $(L,L + dL)$ at time $t$ is given by
\begin{align}
| \Psi_\xi(x = L,t)|^2 \, dL , \label{DofPSQ}
\end{align}
and $x \in \mathbb{R}$. If we take a look at Figure~(\ref{WaveFunction}), we can see that Eqn~(\ref{DofPSQ}) is actually the density of points within this interval at time $t$. Hence, if we consider the entire time interval $[0,T]$ we have
\begin{align}
N^{(S)}_c(L, T; \lambda) = \frac{dL}{T} \int^{T}_0 | \Psi_\xi(L,t)|^2 dt.  \label{SQMPrediction}
\end{align}
Notice that $N^{(S)}_c(L, T; \lambda)$ does not depend on interval constraining the value of the positions. 
\item In the Bohmian interpretation, the number of clicks for the interval $(L,L+dL)$ is actually proportional to the {\it number} of possible initial conditions which can actually hit the detector, 
\begin{align}
\frac{ | \Psi_\xi(x = L,t)|^2 \; dL}{ \int^{q^{max}_0(L + dL)}_{q^{min}_0(L)} dq_0 \; | \Psi_\xi(q(t;q_0),t)|^2},
 \end{align}
where $q(t; q_0)$ is the Bohmian trajectory given in Eqn~(\ref{BS}). If we consider the entire time interval we have
\begin{align}
N^{(B)}_c(L, T; \lambda) = \frac{dL}{T} \int^T_0 dt \, \frac{ | \Psi_\xi(x = L,t)|^2 }{ \int^{q^{max}_0(L + dL)}_{q^{min}_0(L)} dq_0 \; | \Psi_\xi(q(t;q_0),t)|^2}  \label{BQMPrediction}
\end{align}
\end{itemize}
As can be seen, the relations in Eqn~(\ref{SQMPrediction}) and Eqn~(\ref{BQMPrediction}) are rather different from which we can expect different predictions. Being correct this  statement, only an experiment can test will declare which one, $N^{(S)}_c(L, T; \lambda)$ or $N^{(B)}_c(L, T; \lambda)$ matches $N^{exp}_c(L, T; \lambda)$.

A different {\it Gedanken experiment} can be implemented to detect the TOA distribution and the mean TOA. In both cases, the aim should be testing the sensitivity of these quantities to the detector's location as shown in Figures~(\ref{ProbTime}) and (\ref{MeanTimePlot}). In the case of the mean TOA, fixing the squeezing parameters as in Figure~(\ref{MeanTimePlotb}) and considering small values of the ratio $L/l$ seems to be a route worth to explore. The reason is because technically, controlling the squeezing appears to be more difficult than controlling $L/l$.

There are of course major obstacles at the theoretical level and also at towards the experimental realization. In the first case, a reasonable model for the detector's interaction with the squeezed particle is required. Here, only what can be considered the zero order interaction between the particle and the apparatus was analyzed. At the experimental level, we can foreseen a difficulty for implementing small values of the ratio $L/l$ and realistic values of $r$. In both cases these regimes are important or otherwise the time resolution can not separate points and, for example, the curves for large $L/l$ in Figure~(\ref{MeanTimePlotb}) will appear all together.

We consider our results to be the first step towards a deeper analysis linking the squeezed states and the time of arrival measurements. Going further towards higher degrees of freedom systems and also towards considering a model for the detector is the natural next step. We hope, however, that the ubiquitous use of squeezed states in quantum optics experiments might open the doors for an experimental realization confirming or denying any of our predictions or the future to come.

\subsection*{Acknowledgments}
AGCH would like to thank the CONAHCyT for a postdoctoral fellowship.

\bibliographystyle{unsrt}
\bibliography{referencias}

\end{document}